# A Framework for Specific Term Recommendation Systems

Thomas Lüke, Philipp Schaer, Philipp Mayr
GESIS - Leibniz-Institute for the Social Sciences
Unter Sachsenhausen 6-8
50667 Cologne, Germany
{Thomas.Lueke, Philipp.Schaer, Philipp.Mayr}@gesis.org

## ABSTRACT
In this paper we present the IRSA framework that enables the automatic creation of search term suggestion or recommendation systems (TS). Such TS are used to operationalize interactive query expansion and help users in refining their information need in the query formulation phase. Our recent research has shown TS to be more effective when specific to a certain domain. The presented technical framework allows owners of Digital Libraries to create their own specific TS constructed via OAI-harvested metadata with very little effort.

## Categories and Subject Descriptors
H.3.3 [**Information Storage and Retrieval**]: Information Search and Retrieval – Query formulation, Selection process, Search process, Retrieval models

## 1. INTRODUCTION AND MOTIVATION
Modern digital libraries (DL) have grown in size tremendously. While users struggled with empty result sets in the past today's challenges lie within finding the documents most relevant to the users information need. This situation is comparable to the problem in modern web search engines (WSE) where users are confronted with a vast and unknown information space. The well-known vocabulary problem [2] is more prevailing than ever.

To treat this vocabulary problem and other issues during the query formulation phase a wide range of possible query expansion (QE) and search term recommender systems were presented. While in WSE the use of query suggestion systems became omnipresent the situation is different in DL systems (see table 1). Very few DL systems implement interactive query expansion that can be further divided into term suggestions (TS) and query suggestions (QS). In contrast to QS systems that suggest complete query strings TS systems try to add or replace single words or phrases [6]. QS systems are often based on query log analysis but can also be implemented to suggest queries based on the document corpus [1]. In DL systems which are more structured than WSE the use of knowledge organization systems (KOS) is common practice. Typically entire collections in a DL are indexed with controlled terms from a domain-specific KOS like a thesaurus or a classification. The main task of TS systems is to assist users in the process of expressing their information need and supporting them in the formulation of a useful query. These systems try to suggest terms that are closely related both to the users initial query term as well as to the semantic backbone (the KOS) of the DL. While theoretically any kind of metadata may be recommended the promising approach is to recommend terms from KOS.

The curated metadata sets of DL systems are often publicly available via the Open Archives Initiative Protocol for Metadata Harvesting (OAI-PMH[1]) interface. OAI-PMH is a "low-barrier" quasi-standard and specification for repository interoperability with a very limited set of HTTP services. Structured metadata can be easily exposed via OAI-PMH. Because most DL and document repository systems support OAI-PMH our framework builds on this specification.

Search term suggestion has been proven to be helpful in real-world DL systems [3] as well as in standard IR retrieval tests (using test corpora like GIRT or iSearch). Our recent research on TS has especially proven domain specific recommendations to give better performance compared to general recommendations based on query logs. In [4] we created 17 TS systems from the social sciences domain, 16 based on specific sub-disciplines and one giving general recommendations. We compared the retrieval performance of these TS. The main findings show that automatic QE with specific TS leads to significantly better results than QE with a general TS.

**Table 1: Digital library systems and academic WSE and their ability to offer term suggestion (TS), query suggestion (QS) and structured metadata using the OAI-PMH interface.**

| Site | TS | QS | OAI-PMH |
| --- | --- | --- | --- |
| ACM Digital Library | no | no | no |
| Google Scholar | no | no | no |
| MIT Repository | no | no | yes |
| arXiv | no | no | yes |
| PubMed | yes | no | yes |
| MS Academic Search | yes | yes | no |

Most DL systems lack any kind of query or term suggestion mechanism although they include all the necessary structured data to implement such a system (see some examples in table 1). In contrast WSE-based systems like Microsoft Academic Search do include such interactive query expansion mechanisms.

---

[1] http://www.openarchives.org/OAI/openarchivesprotocol.html

We therefore propose a web-based interactive system to easily construct custom TS systems that use available corpus information harvested through a DL's OAI-PMH interface. Our system calculates the semantic relatedness between title and abstract terms (the "free terms") and the controlled terms for the entire document corpus. Using this approach we can offer user-defined KOS-based term suggestions for every "free" query term. These TS systems are specific for every single DL registered at IRSA.

As an example users who are looking for the string "youth unemployment" in a social sciences context the TS system will provide search term suggestions like "labor market" or "education measure" that are semantically related to the initial query. Another possible suggestion might be "adolescent", which is a controlled term for "youth". The suggestions of our TS approach go far beyond simple term completion [5] and can support the search experience as we have shown in [3].

## 2. THE IRSA FRAMEWORK

IRSA is accessible through a web frontend that handles user registration, status mails and also provides rudimentary management and accessibility methods for the domain specific TS. To make use of the framework DL operators have to take six steps: (1) Register at the project's website[2], (2) create a new repository within the system, (3) supply either the OAI-PMH interface URL of the DL or the metadata itself as XML files in oai_dc style[3], (4) schedule the repository for processing (5), wait until processing of the TS is finished (typically within hours), and (6) use the generated RESTful web service in the specific project. This web service can be included in the DL with a few lines of code depending on the programming language and frameworks used. An API key is used to ensure privacy for each user. Figure 1 illustrates the internal workflow with most steps happening without user interaction.

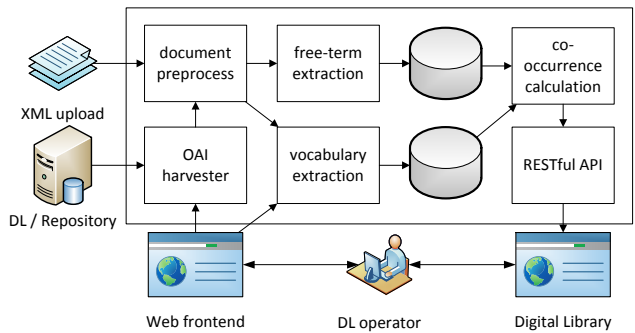

**Figure 1: Workflow of the IRSA system.**

After a user has scheduled the processing of a new repository the framework starts harvesting the data. When the entire repository has been harvested processing of the TS will start automatically. The system counts the co-occurrences of terms in a set of given metadata fields. By default terms from a documents title (dc:title) and abstract (dc:description) are matched against co-occurring terms from the controlled vocabulary (dc:subject) found as subjects of a document. Although any two metadata fields may be used to calculate recommendations, the use of qualified Dublin Core is advised as it allows further classification of metadata, i.e. providing language information. IRSA can not only create recommendations of subjects or terms but also other fields like journals or years of publication (i.e. to detect trends for certain topics in a DL). All data is saved into a PostgreSQL database where some of the co-occurrence calculations are done. Web frontend, data-harvesting & -parsing and user management are part of software written in Grails technology. The entire framework is open source software and may therefore be set up in environments where the use of an external web service might be unwanted.

---

[2] http://www.gesis.org/irm

[3] http://www.openarchives.org/OAI/2.0/oai_dc.xsd

Internally the co-occurrences are calculated using the Jaccard index where term y is considered most related to another term x (x and y from different metadata fields) if J(x,y) is higher than J(x,z) for any term z which is unequal to y:

$$J(x, y) = \frac{|DS_x \cap DS_y|}{|DS_x \cup DS_y|} = \frac{df_{xy}}{df_x + df_y - df_{xy}} \quad (1)$$

Recommendations for term x are based on the terms most related to x in descending order.

The Jaccard index was chosen as a robust alternative to a more complex solution using Support Vector Machines (SVM) and Probabilistic Latent Semantic Indexing (PLSA) used in [3]. Besides delivering the better retrieval performances the implementation based on the Jaccard index is significantly faster. While the SVM and PLSA implementations needed 3 days to calculate the co-occurrences of ~400.000 documents, the new implementation can do the same in just 2 hours. Our evaluation showed that the simple Jaccard index still delivers results that are, in the use case of term recommendation, comparable or better than the tested PLSA/SVM approach. Other metrics like the Normalized Web Distance or the Dice coefficient could be used for calculation.

In this paper we presented a "low barrier" approach to build specific KOS-based term suggestion systems for DL that are OAI-PMH compatible. The IRSA framework reduces the complexity of creating and hosting such TS systems and the resulting web services can easily be integrated into existing DL.

## 3. ACKNOWLEDGEMENTS
This work was funded by DFG, grant no. SU 647/5-2.